\documentclass[twocolumn,aps,prl,showpacs,superscriptaddress]{revtex4-1}

\usepackage{graphicx,color,hyperref}

\usepackage{amsmath}
\usepackage{amsfonts}
\usepackage{amssymb}
\usepackage{dsfont}

\begin{document}

\title{SYK non Fermi Liquid Correlations in  Nanoscopic Quantum Transport  
}

 \author{Alexander Altland}

\affiliation{Institut f\"ur Theoretische Physik, Universit\"at zu K\"oln,
Z\"ulpicher Stra\ss e 77, 50937 K\"oln, Germany}

\author{Dmitry Bagrets}

\affiliation{Institut f\"ur Theoretische Physik, Universit\"at zu K\"oln,
Z\"ulpicher Stra\ss e 77, 50937 K\"oln, Germany}

\author{Alex Kamenev}

\affiliation{W. I. Fine Theoretical Physics Institute and School of Physics and Astronomy, University
of Minnesota, Minneapolis, MN 55455, USA}

\date{\today}


\begin{abstract}
Electronic transport in nano-structures, such as long molecules or 2D exfoliated flakes, often  goes through a nearly degenerate set of single-particle orbitals. Here we show that in such  cases a conspiracy of the narrow band and strong {\em e-e} interactions may stabilize a non Fermi liquid phase in the universality class of the complex Sachdev-Ye-Kitaev (SYK) model. Focusing on signatures in quantum transport, we demonstrate the existence of  anomalous power laws in the temperature dependent conductance, including algebraic scaling $T^{3/2}$  in the inelastic cotunneling channel, separated from the conventional Fermi liquid $T^2$ scaling via a quantum phase transition. The relatively robust conditions under which these results are obtained indicate that the SYK non Fermi liquid universality class might be not as exotic as previously thought.
\end{abstract}

\maketitle

\noindent \emph{Introduction:} Electronic device miniaturization is now routinely
operating at levels where quantum limits are reached. Examples where quantum effects
are of key relevance and/or used as operational resources include single molecule
transport\cite{Vilan2017,Merces2017,Su2016,Porath2005}, various realizations of qubits\cite{Kouwenhoven2002}, and increasingly even
commercial applications such as Q-dot display technology. Physically, such nanoscopic
devices (henceforth summarily denoted as `quantum dots') are frequently described\cite{Nazarov2009} in
terms of only few collective variables --- their cumulative electric charge, a
global superconducting order parameter, a collective spin, etc.

Starting from  first principle many body representations, this `universal Hamiltonian' 
approach\cite{Kurland2000,Aleiner2002} is
implemented via elimination of microscopic degrees of freedom which in turn rests 
on statistical arguments\cite{Kurland2000,Alhassid2000,Urbina2013}. To illustrate the principle in the simplest physical
setting, consider  a small quantum system with $i=1,\dots, N$ electronic orbitals, assumed spinless for simplicity. 
This setting is described by the Hamiltonian  $\hat H =\sum_{i}^N \epsilon_i c^\dagger_i  c_i  
    + \sum_{ijkl}^N \tilde J_{ijkl}c^\dagger_i c^\dagger_j c_k c_l$,
      where $\epsilon_i$ are the energies of the non-interacting orbitals, and
      $\tilde J_{ijkl}$ are the matrix elements of the particle interactions ---
      generally strong in the case of nanoscopic device extensions. Systems of realistic
      complexity are typically  non-integrable on the single particle level, implying
      effectively random matrix elements, $\tilde J_{ijkl}$. This randomness is
      usually taken as justification to discard all matrix elements except those with
      non-zero mean value. Specifically, focusing  on contributions with $i=k$,
      $j=l$, or $i=l$, $j=k$, and assuming approximate equality of diagonal matrix
      elements on average, one is led to the representation
\begin{align}
\label{eq:SYKBasic}
    \hat H =  \sum_{i}^N \epsilon_i c^\dagger_i  c_i + \tfrac 12  E_C\, \hat n^2 
    + \sum\limits_{ijkl}^N J_{ijkl}c^\dagger_i c^\dagger_j c_k c_l,
\end{align}
where $J_{ijkl}$ now excludes matrix elements with identical indices, 
$\hat n=\sum_i^N c^\dagger_i c_i$ is the total charge on the dot and the coefficient 
$E_C=e^2/C$ defines its effective electrostatic capacitance, $C$. The standard universal
Hamiltonian approach~\cite{Kurland2000,Aleiner2002} defines $\hat n$ as the central collective variable, and
ignores the contribution of the random sign matrix elements, $J_{ijlk}$, to the interaction energy.

In this paper, we caution that the neglect of the term $\hat H_\mathrm{SYK}\equiv
\sum J_{ijkl}c^\dagger_i c^\dagger_j c_k c_l$ may be less innocent then
is commonly assumed. The point is that $\hat H_\mathrm{SYK}$ is a variant of the
complex SYK Hamiltonian~\cite{Sachdev-Ye,kitaev2015talk}\footnote{The difference is
that in the standard SYK-Hamiltonian the couplings are drawn from a Gaussian
distribution. However, in view of the effectively Gaussian distribution of chaotic
single particle wave functions, this difference is inessential.}, the latter being
defined as an all--to--all interaction Hamiltonian with random matrix elements taken
from a zero mean Gaussian distribution with $\langle J_{ijkl}^2\rangle=J^2/N^3$. The
pure SYK Hamiltonian
\cite{Sachdev-Ye,kitaev2015talk,davison2017thermoelectric,fu2016numerical,Garcia16,sonner2017eigenstate,Garcia17,altland2018quantum}
defines a universality class distinguished for  a maximal level of entanglement,
chaos, and non Fermi liquid (NFL) correlations, otherwise shown only by black holes
(in 2D gravity the latter are related to SYK model via the holographic correspondence
\cite{CommentsSYK16,cotler2017black,engelsoy2016investigation}). Correlations
generated by arrays of SYK cells are increasingly
believed\cite{Balents2017,berkooz2017higher,bi2017instability,gu2017local,Altland2019}
to be  relevant in the physics of strongly correlated quantum matter, and it has been
suggested that single copies of SYK-Hamiltonians might describe small sized samples
of flat band materials\cite{Franz2019}. In the following, we reason that even the low
temperature physics of the much more generic class of systems described by the
Hamiltonian above can be partially, or even fully governed by the SYK universality
class. The latter is the case if the band width, $W$, of single particle orbitals,
$\epsilon_i$, is smaller than the interaction strength, $W<J$. For these values,
strong quantum fluctuations generated by $\hat H_\mathrm{SYK}$ render the single
particle contribution $\hat H_0\equiv \sum_i \epsilon_i c^\dagger_i c_i$
irrelevant\cite{Feigelman2018}. Conversely, for larger values the fluctuations
themselves get suppressed by $\hat H_0$. However, even then the presence of $\hat
H_\mathrm{SYK}$ shows in extended crossover windows in temperature where the dot
shows NFL correlations. In the following we address both cases,  focusing on
signatures on low temperature transport.

The crucial feature of the SYK Hamiltonian is the presence of a weakly broken infinite dimensional conformal symmetry
\cite{kitaev2015talk,Sachdev,Maldacena16}. This
symmetry breaking manifests itself in NFL correlations, and in the emergence of a
set of Goldstone modes, which in the present context define a second set of low
energy collective variables, $h(\tau)$, besides $\hat n(\tau)$. Depending on temperature, and the relative strength of
interactions and the single-particle bandwidth, the conspiracy of these degrees of freedom can drive the system into a strongly correlated NFL phase of
matter. In quantum transport, the presence of these regimes shows in non-monotonicity of the temperature dependent conductance, $g(T)$, and in power laws  $g(T)\sim T^\alpha$ different from the $T^2$ of the Fermi liquid dot.

\noindent \emph{Symmetries:} We start by identifying the symmetries of the system, which in turn determine
its low-energy quantum fluctuations. This is best done in a coherent state
representation, where the Hamiltonian is expressed via Grassmann valued
time-dependent fields $ (c,c^\dagger)\to (c(\tau), \bar c(\tau))$, depending on
imaginary time $\tau\in [0,\beta]$. The system's action $S=\int d\tau (\bar c_i
\partial_\tau c_i -H(c,\bar c))$  is then approximately invariant under the
transformations\cite{kitaev2015talk,Sachdev,Maldacena16} 
\begin{equation}
    \label{FieldTransformation}
    c_i(\tau) \rightarrow 
e^{-i\phi(\tau)} 
 \left[ \dot h(\tau) \right]^{1/4} c_i(h(\tau)), 
\end{equation}
where the dot stands for the time derivative, and the $\mathrm{U}(1)$ phase  $\phi$ is canonically conjugated to
the charge operator, $\hat n$. The functions $h(\tau)$ are diffeomorphic
reparameterizations of imaginary time and as such take values in the coset space
$\mathrm{Diff}(S^1)/\mathrm{SL}(2,R)$, where $\mathrm{Diff}(S^1)$ is the set of smooth functions parameterizing the periodic interval of imaginary time and the  factorization of $\mathrm{SL}(2,R)$
accounts for  a few exact global symmetries of the action). The symmetries (\ref{FieldTransformation}) are explicitly broken  by both, the time derivative in the action, and the single particle contribution in Eq.~(\ref{eq:SYKBasic}). We first discuss the former and note that the action cost associated with temporal fluctuations of $(\phi,h)$ reads\cite{kitaev2015talk,Sachdev,Maldacena16} 
\begin{align}
    \label{Schwarzian}
     S_0[\phi,h] \!= \!\!\int\!\! d\tau\! \left[ \tfrac 12 E_C^{-1} \dot\phi^2 - m\{h,\tau\}\right],  
     \end{align}   
where 
$\{h,\tau\}\equiv  (h''/h')' - \frac 12 (h''/h')^2$ is the Schwarzian derivative, 
the mass $m\propto  N/J$ and we assume the following hierarchy of energy scales: $ m^{-1} \ll  E_c \ll  J$\footnote{Assuming all matrix elements being comparable, $J_{ijkl}\approx E_c$, one finds $J\approx N^{3/2}E_c\gg E_c$}.

The  physical information relevant to our discussion below is contained in the fermion Green functions, $G_{\tau_1,\tau_2}=\langle c_i(\tau_1)\bar c_i(\tau_2)\rangle$. The transformations (\ref{FieldTransformation})  affects the Green functions as $G\to G[\phi,h]$, where  
\begin{align}
					\label{GreenFunction}
&G_{\tau_1,\tau_2}[\phi,h] = e^{-i\phi(\tau_1)} G_{\tau_1,\tau_2}[h] \, e^{i\phi(\tau_2)};  \\
&G_{\tau_1,\tau_2}[h] =  - \mathrm{sign}(\tau_1-\tau_2) \, 
\left[ \frac{\dot h(\tau_1)   \dot h(\tau_2)} {( h(\tau_1) -   h(\tau_2))^2} \right]^{1/4}.   \nonumber  
\end{align}
Here,  the terms in the numerator are consequences of
Eq.~\eqref{FieldTransformation}, and the denominator reflects the `engineering
dimension' $\Delta=1/4$ of fermions in a mean field approach to the SYK Hamiltonian\cite{Sachdev,Kitaev}. In the absence of
reparameterizations, this leads to the NFL scaling $G_{\tau_1,\tau_2}\sim |\tau_1-\tau_2|^{-1/2}$ signifying an interaction--dominated theory.

\begin{figure}
  \includegraphics[width=9cm]{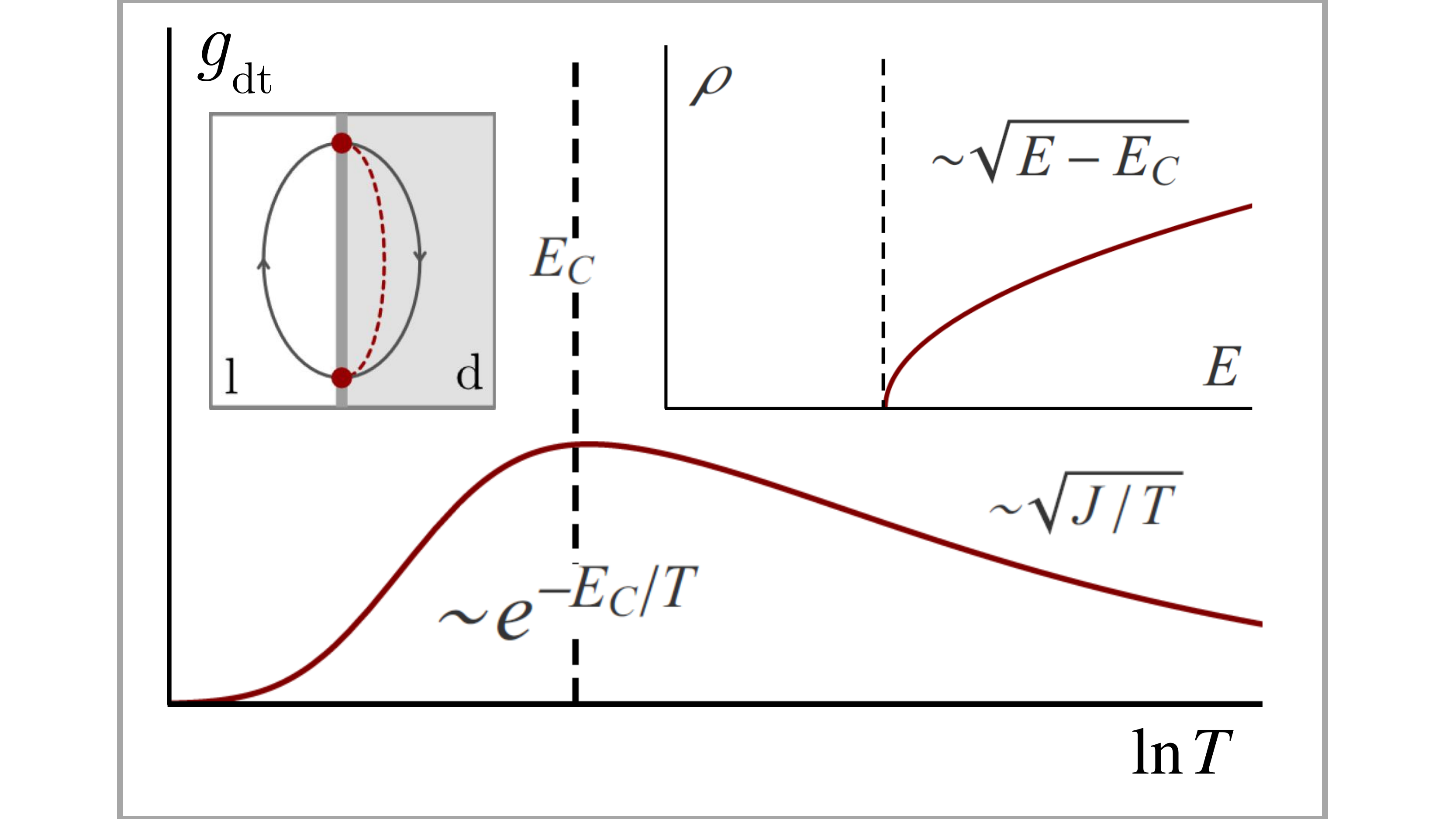}
  \caption{Main Panel: temperature dependence of the direct tunneling contribution to
  the conductance. An exponential suppression at temperatures $T<E_C$ gives way to an
  $T^{-1/2}$ power law at larger temperatures. Left inset: diagram of the direct
  tunneling process;  black lines are Green functions in the lead (l) and the dot
  (d),  fat dots are tunneling vertexes  at times $\tau_{1,2}$, respectively, and the
  red dashed line represents the charging correlation $D(\tau_1-\tau_2)$,
  Eq.~(\ref{CB}). Right inset: energy dependence of the average tunneling density of
  states on the dot at $T=0$.}
  \label{fig:TunnelingConductance}
\end{figure}

\noindent \emph{Isolated dot:} For an isolated dot integration over the phase field
with the action~\eqref{W-action} generates the factor\cite{Kamenev1996,Wang1996}
\begin{equation}
				\label{CB}
D(\tau_1-\tau_2)\equiv \left\langle e^{-i\phi(\tau_1)} e^{i\phi(\tau_2)}\right\rangle_\phi = e^{-E_C |\tau_1-\tau_2|/2}.
\end{equation}
Fourier transformation of $D(\tau)$ to the energy domain leads to a gap, $E_C$, in the excitation spectrum, and an exponential suppression of the 
 single-particle density of states, the Coulomb blockade. The second factor $\left\langle G_{\tau_1,\tau_2}[h]
\right\rangle_h$, which in the same from appears in the charge-neutral Majorana SYK
model, has been studied extensively in Refs.~\cite{Bagrets-Altland-Kamenev2016, Mertens2017}. Here, the main observation
is a crossover from temporal decay as $|\tau_1-\tau_2|^{-1/2}$ for intermediate time
scales $ 1<J|\tau_1-\tau_2|<N$, 
to  $|\tau_1-\tau_2|^{-3/2}$ in the fluctuation
dominated long time regime,  $N<J|\tau_1-\tau_2|$. This change implies a crossover
in  the effective fermion dimension from $\Delta=1/4$ to $\Delta=3/4$. Fourier transformation of the long time power law reveals the presence of a soft zero-bias anomaly $\propto \sqrt{E-E_c}$ on top
of the hard Coulomb blockade gap (see inset in Fig.~\ref{fig:TunnelingConductance}.)

\noindent \emph{Tunneling conductance: } We next consider  the system   connected to metallic
leads via the tunneling Hamiltonian $H_T=\sum_{i,k}
V_{ik} c^\dagger_i d_k+h.c.$. Here, $d_k$ are annihilation operators in the normal
leads and we assume the matrix elements $V_{ik}$ to be effectively random with variance $\langle |V_{ik}|^2\rangle \equiv v^2$.  To  second order in perturbation theory, this generates the tunneling action
\begin{align}
    \label{AES}
     S_T[\phi,h] \!= - g_0 T\!\!\int\!\!\!\!\int\!\!  d^2\tau
     \frac{e^{ -i \phi(\tau_2)} G_{\tau_2,\tau_1}[h] e^{  i \phi(\tau_1)}}{\sin( \pi T(\tau_1-\tau_2))}\,   ,  
     \end{align}  
where $g_0\propto  \nu v^2 N/J $ is the bare dimensionless tunneling coupling of the lead-dot
interface  and $\nu$ is the DoS in the leads. 
Equation~\eqref{AES} generalizes the celebrated Ambegaokar, Eckern and
Sch\"on (AES) \cite{AES} action for metallic quantum dots to system with NFL
correlations. The difference amounts to the generalization $ (\tau_1-\tau_2)^{-1}\to G_{\tau_1,\tau_2}[h]$, and this modified AES approach defines our starting point for the description of low
temperature quantum transport.

In the following, we focus on the temperature
dependent  conductance, $g=g(T)$, as an observable diagnosing the presence of NFL
correlations via anomalous power laws. The linear dc conductance through the dot is conveniently obtained by differentiation  $g\sim \lim_{\omega\to 0}\omega^{-1}\delta^2 \ln Z[{\cal A}]/\delta {\cal A}_\omega \delta {\cal
A}_{-\omega}|_{{\cal A}=0}$, of the generating function  $Z[{\cal A}]=\int {\cal D}\phi {\cal D}h
\exp(-S_0[\phi,h]-S_T[\phi+{\cal A},h])$ in a source vector potential ${\cal
A}(\tau)$, minimally coupled to the tunneling action.

To the first order in the tunneling conductance $g_0\ll 1$ \footnote{Since $g_0\propto v^2N$, in the limit of large number of channels, $N$, the matrix elements, $v$, should decrease to keep $g_0~\ll~1$.}  the calculation of $g(T)$ reduces to
that   of the Green function, $\left\langle G[\phi,h] \right\rangle_{\phi,h}$,  of the
isolated dot. In this approximation, the conductance $g\simeq g_\mathrm{dt}$ probes the {\em direct tunneling} processes amplitude converting a lead quasiparticle into a single-particle excitation of the dot (see the left inset in Fig.~\ref{fig:TunnelingConductance}).  With the 
time dependence of $\langle G_{\tau_1,\tau_2}[\phi,h]\rangle_{\phi,h}$, stated above, one then obtains (cf. Fig.~\ref{fig:TunnelingConductance})
\begin{equation}
					\label{first-order}
g_\mathrm{dt}(T) \propto g_0 \left\{ \begin{array}{ll} 
 e^{-E_C/T}; &\quad T<E_C,\\
\sqrt{J/T}; & \quad E_C<T<J,
\end{array}
\right.					
\end{equation}
where the first line  is an immediate consequence of the Coulomb gap,
Eq.~(\ref{CB}). The second one follows from the smallness of fluctuations of both $\phi$ and $h$ in the high temperature
regime $T>E_C>J/N$, implying that the Green function $\left\langle G_{\tau_1,\tau_2}[\phi,h]
\right\rangle_{\phi,h}\sim |\tau_1-\tau_2|^{-1/2}$ assumes its mean field NFL form\cite{Gnezdilov2018,Franz2019}. This leads to the non-monotonous
temperature dependence of the direct tunneling conductance, as shown in Fig.~\ref{fig:TunnelingConductance}.

\begin{figure}
  \includegraphics[width=9cm]{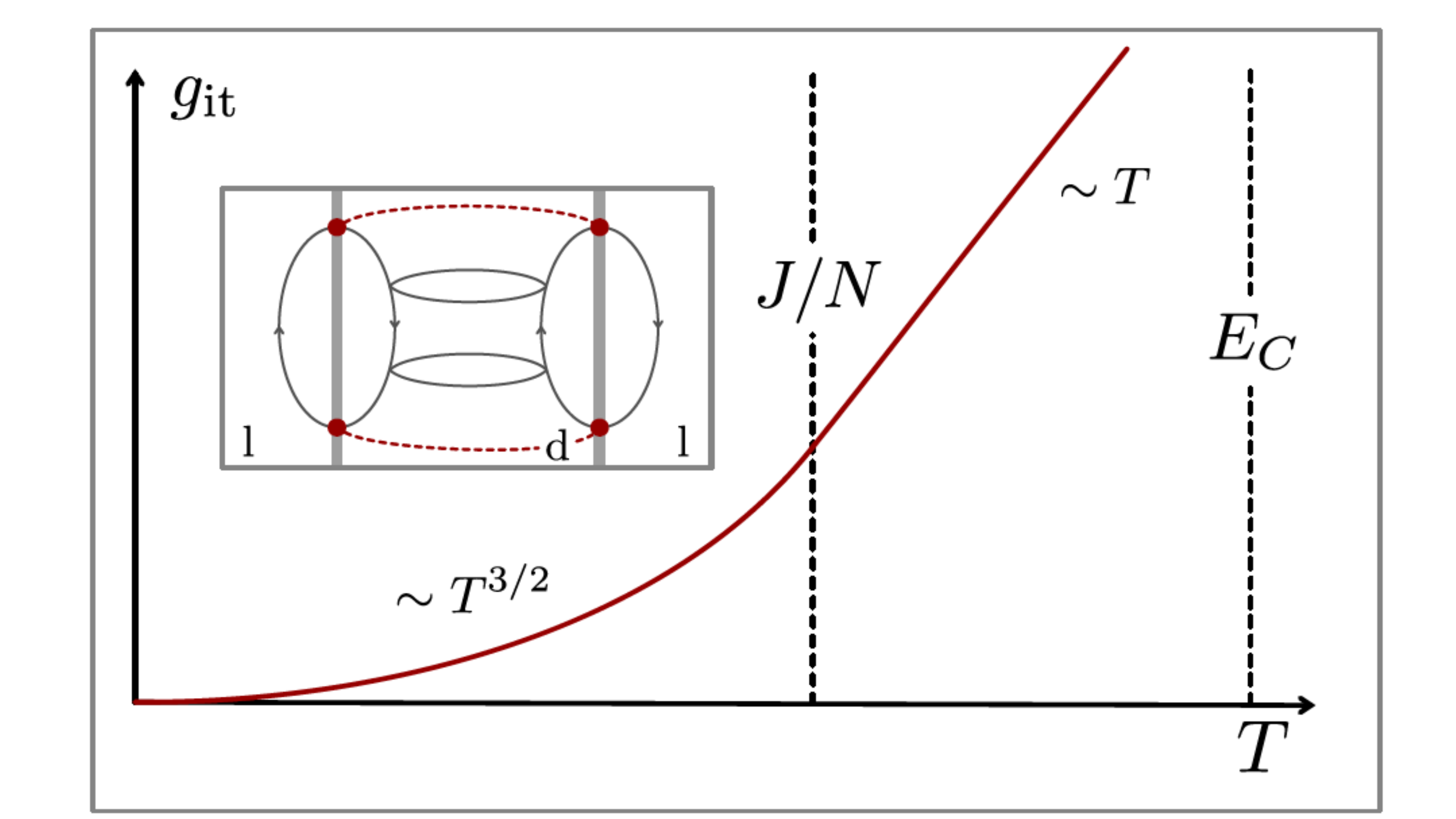}
  \caption{Main Panel: schematic temperature dependence of the inelastic cotunneling
  contribution to the conductance.  Inset: Inelastic cotunneling diagram, where a
  particle enters the dot from the left lead, while another particle exits to the right a
  short time $\sim E_C^{-1}$ after, and NFL `particle-hole' excitation with the energy $\sim T$ is left behind on
  the dot.}
  \label{fig:InelasticCotunneling}
\end{figure}

\noindent \emph{Inelastic cotunneling:} At low temperatures, $T<E_C$, the direct
tunneling transport is taken over by a  process of  second order in $g_0$, which
escapes exponential suppression. This transport channel,   colloquially known as {\em
inelastic cotunneling} \cite{AverinNazarov,Aleiner2002},
$g_\mathrm{it}$, is a cooperative process where the tunneling of an electron onto the
dot  creates a virtual state, which  relaxes after a short time, $\sim
E_C^{-1}$, via the exit of  {\em another} electron into the other lead. For a
metallic dot, the corresponding contribution to the conductance reads
$g_\mathrm{it}(T)=g_0^2 T^2/E_C^2$    \cite{AverinNazarov}, where the factor $T^2$  measures
the phase volume available for particle-hole creation, and $E_C^{-2}$ is the energy
denominator picked up during the virtual excitation of the dot. A calculation
outlined in the Supplemental Material shows~\footnote{See Supplemental Material for the
technical derivations of the results presented in the main text.
} 
that for the NFL dot, this result changes to (cf. Fig.~\ref{fig:InelasticCotunneling})
\begin{equation}
					\label{second-order}
g_\mathrm{it}(T)=\frac{g_0^2}{E_C^2}\, \left\{ \begin{array}{ll} 
\sqrt{NJ}\, T^{3/2}; &\quad T<J/N,\\
JT;    & \quad  J/N<T<E_C, 
\end{array}
\right.					
\end{equation}
where the power law at intermediate temperatures reflects the modified temporal exponent $(\tau^{-1})^2\to (\tau^{-1/2})^2$ for `particle-hole' propagation as described by the NFL mean-field Green function. At lowest temperatures, $T<J/N$, strong reparameterization  fluctuations result \cite{Bagrets-Altland-Kamenev2016} in the `particle-hole' propagator in the dot of the form $\langle (G_\tau[h])^2\rangle_h\propto \tau^{-3/2}$, leading to an unconventional fractional exponent $3/2$. 

Equations~\eqref{first-order} and~\eqref{second-order} are our main results for the signatures of NFL correlations in quantum transport. Compared to a metallic system, the existence of the fluctuation scale $J/N$ implies a higher amount of structure, i.e. different power laws, and regimes of non-monotonic temperature dependence.  In the following, we ask how stable these findings are against perturbations, notably the presence of the free electron contribution to realistic model Hamiltonians.

\noindent \emph{Stability of the NFL phase:} Turning to the role played by the single particle Hamiltonian, $\hat{H_0}$, we first
consider the case of high temperatures $T>J/N$, where fluctuations of the conformal
Goldstone modes are inessential. In the absence of $\hat{H}_0$, single particle
excitations, then propagate via $G\sim (J T)^{-1/2}$, reflecting the NLF particle
dimension, $\Delta=1/4$.  Second order perturbation theory in $\hat{H}_0$ leads to
the energy shift $W^2G\sim W^2/(JT)^{1/2}$, proportional to the square of the single
particle bandwidth. This shift should be compared with the self consistent self
energy due to interactions, $\sim J^2G^3\sim J^2/(JT)^{3/2}$\cite{Balents2017}. Equating these two
scales, we find that for temperatures lower than $T\sim W^2/J$, $\hat H_0$ dominates
and the dot turns into  a FL subject to capacitive interactions. For larger
temperatures, between $W^2/J$ and the high energy scale of the SYK Hamiltonian, $J$,
it is in a NFL regime. This window exists provided that $J>W$, which sets a
zeroth order criterion for the observability of transport signatures such as such as
the second line of~\eqref{second-order}.

Turning to the more subtle case of low temperatures, $T<J/N$, we reason that in this
case there exists a critical value $W_c\propto J/N$ below which the low temperature
physics is governed by strong NFL interactions. Conversely, larger band widths
stabilize the FL dot. The Existence of this phase transition was first noted by 
Lunkin, Tikhonov and Feigelman\cite{Feigelman2018}. Paraphrasing their arguments,  large $W$ suppresses reparametrization fluctuations, 
enforcing the mean-field NFL scaling dimension $\Delta=1/4$. In this case, the single particle term $\sim \int d\tau \bar c_i c_i$ 
carries dimension $\tau^{1-2\Delta}=\tau^{1/2}$, and hence is relevant in a 
renormalization group (RG) sense. Conversely, if the single particle term is initially weak, 
fluctuations induce a dimensional crossover $\Delta\to 3/4$. In this case, the dimension of the single particle term is $\tau^{-1/2}$, indicating irrelevancy. The two scenarios must be separated by a
critical value, $W_c$, which, however, this simple argument  is not
able to predict.

To better understand the transition, we develop its RG treatment along the lines of
Ref.~\cite{Altland2019}, where Majorana SYK arrays were considered.  We begin by integrating over the fermion fields and averaging the
action over a random distribution of $\epsilon_i$ to generate the term
\begin{align}
    \label{W-action}
     S_W[h] \!= w \!\!\int\!\!\!\!\int\!\!  d^2\tau \, G_{\tau_1,\tau_2}[h]G_{\tau_2,\tau_1}[h],  
     \end{align}  
where  $w=NW^2/J$. Note that the number conservation  of the
single particle Hamiltonian implies that $S_W$ depends on the field $h(\tau)$, but not on
$\phi(\tau)$. Since the tunneling probe action~\eqref{AES} can be assumed arbitrarily weak we need to consider the two running constants $m$ and $w$ of the internal actions Eqs.~\eqref{Schwarzian} and
\eqref{W-action}, respectively. The
renormalizability of this theory is safeguarded  by its exact $\mathrm{SL}(2,R)$
symmetry, which constrains the form of relevant contributions to the action. In the supplementary material we show that the flow of the two couplings is governed by the equations
\begin{equation}
 						\label{eq:RG-1order}
\frac{d\ln m}{dl} =-1 + \frac{1}{24} w m;\quad\quad \frac{d\ln w}{dl} = {1\over 2}.
\end{equation} 
These equations are best analyzed by defining $\lambda \equiv wm$, which leads to the closed equation  $d\ln \lambda/dl = \lambda/24-1/2$. This indicates a critical value $\lambda_c=12$, separating a NFL phase at smaller $\lambda$ from  FL phase at larger $\lambda$.  In terms of the bare parameters $\lambda \propto (NW/J)^2$ and thus $W_c\propto J/N$.

\begin{figure}
  \includegraphics[width=9cm]{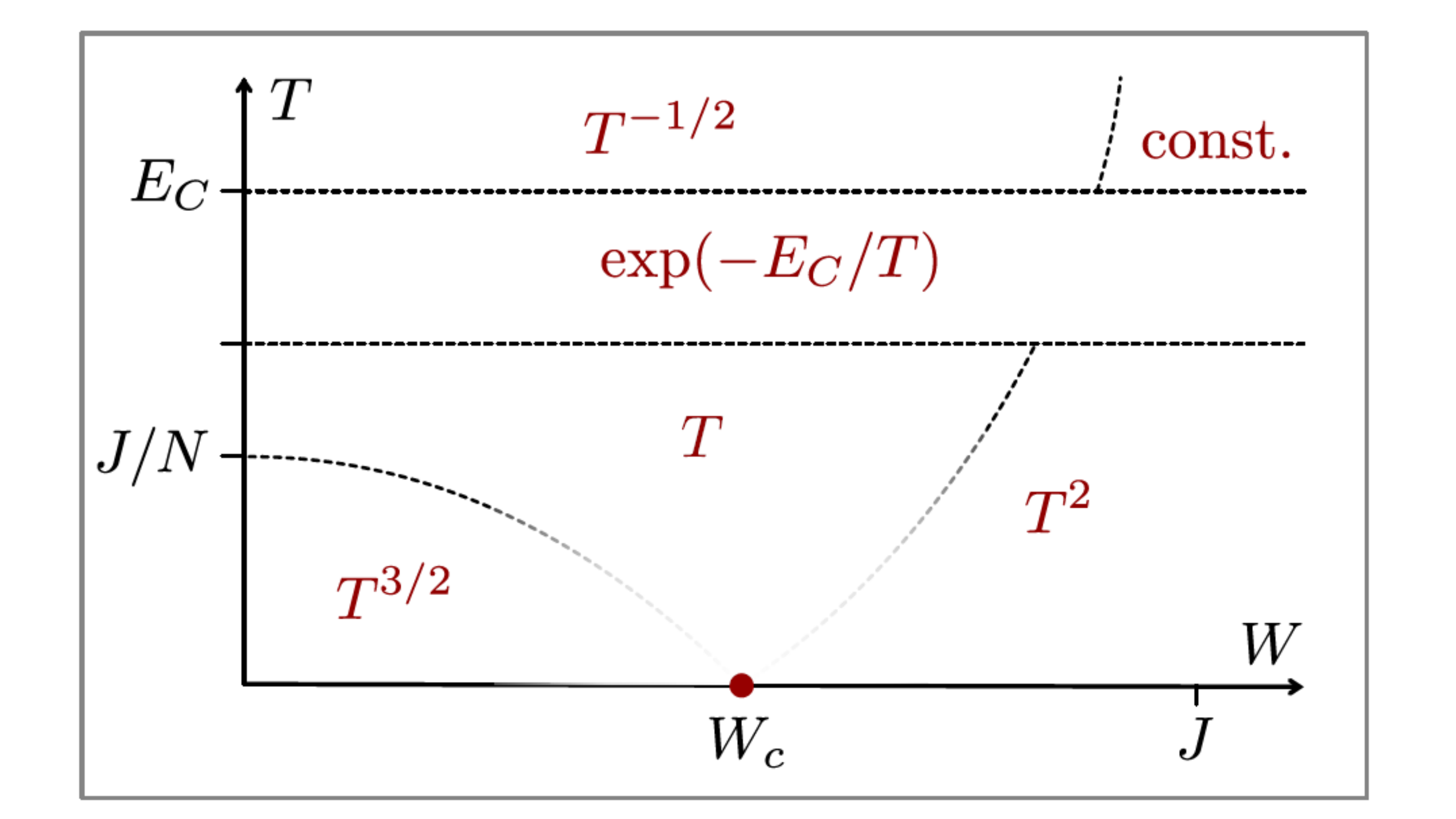}
  \caption{Summary of the different regimes characterized by algebraic or exponential temperature dependence of the linear conductance through a quantum device with a narrow single particle band, $W$. }
  \label{fig:PhaseDiagram}
\end{figure}

\noindent\emph{Phase diagram:} A summary of the regimes with different algebraic or
exponential temperature dependence of the conductance is shown in
Fig.~\ref{fig:PhaseDiagram}. The two essential parameters organizing these regimes
are the band width, $W$, and  temperature, $T$. At zero temperature, a
quantum critical point at $W_c\propto J/N$ separates a wide band width Fermi liquid phase renormalized by charging effects from
a complementary  NFL phase  governed by the SYK Hamiltonian. The two
domains are separated by a  region where the temperature dependence of the
conductance and of other observables is determined by the critical exponents of the
phase transition. A detailed analysis of this regime is beyond the scope of the
present paper. However, off criticality, we expect the formation of robust power
laws, provided the dynamics on the single particle level is sufficiently chaotic, the
number of involved orbitals  is not too large ($W_c\propto N^{-1}$), and interactions
are strong.

\noindent\emph{Summary and discussion:} In this paper, we have drawn a bridge between
the physics of low capacitance quantum devices and that of the SYK
Hamiltonian. The observation that signatures of the SYK universality class might  show in a wider class of systems rests on two principles: first,
the conspiracy of interactions and chaoticity of single-particle wave
functions  makes the complex SYK Hamiltonian  a natural contribution
in the  description of nanoscopic quantum systems. Second, at low energies, we then see the emergence of two soft modes, one,
$\phi(\tau)$, representing soft fluctuations of the $\mathrm{U}(1)$ charge mode, and
another, $h(\tau)$, that of conformal symmetry breaking. (The sole difference between
the complex SYK system and the Majorana version, featuring in connection with
holography, is the presence of the former mode.) As exemplified above, the
respective effects of these two modes can be largely separated in the description of
physical observables. Specifically, we have seen that low temperature quantum
transport is strongly influenced by the infrared physics of the reparameterization
mode. At the same time, we have also seen that the observability of these effects
requires relatively narrow single-particle bands. Candidate systems  where the effects discussed above might become observable include complex molecules\cite{Merces2017}, `artificial atoms' based on semiconductor platforms, or exfoliated 2D materials. In view of the results discussed above it would be intriguing to search for signatures of NFL physics in such systems.
 
{\it Acknowledgements ---} 
AA and DB was funded  by the Deutsche Forschungsgemeinschaft (DFG)
Projektnummer 277101999 TRR~183 (project A03).  A.K. was supported by NSF grant DMR-1608238.

\newpage
{\bf Supplementary material}
\vskip 0.1cm

\emph{Derivation of the $S_0[\phi,h]$ action:}  
We here sketch how the action~(\ref{Schwarzian}) follows from an expansion of the theory in the derivative operator $\hat \partial_\tau$. Our starting point is the fermion determinant 
\begin{equation}
\label{eq:S_f_phi}
S[\phi,h] = -N  {\rm Tr}\ln\left( 1 - \hat \partial_\tau G[\phi,h] \right)
\end{equation}
appearing as part of the exact action describing the disorder averaged SYK partition sum (c.f.~Appendix A of
Ref.~\cite{Bagrets-Altland-Kamenev2016}). The action $S[\phi,h]$ contains the fluctuation-dressed
single particle Green function in combination with the time derivative operator which breaks the
symmetry defined by Eq.~(\ref{FieldTransformation}). We are going to demonstrate that to second
order in a gradient expansion the effective action assumes the form
\begin{equation}
\label{eq:S_loc_SY}
S_2[h,\phi] = \int d\tau \left[ \tfrac 12 K \dot \phi^2 - m \{ h,\tau  \}  - i {\cal L}\, \dot \sigma \, \dot \phi \right] ,
\end{equation}
where $\sigma(\tau) \equiv \ln \dot h(\tau)$ is the non-compact degree of freedom of Liouvillian quantum mechanics~\cite{Bagrets-Altland-Kamenev2016} and the coupling constants $K = 32 m$ and ${\cal L} = K/(4 \ln N)$, contain the Liouville `mass' with bare value
\begin{equation}
\label{eq:m}
m=  \frac{N \ln N }{64 J}\, \sqrt{ \frac{\cos 2\theta }{{2\pi}}}.
\end{equation}
Here, the parameter $\theta$ is determined by the average occupation number of the dot
$Q$ as~\cite{Sachdev}
\begin{equation}
Q \equiv \frac{\langle \hat n\rangle} {N} = \frac 12  - \frac \theta \pi - \frac{\sin(2\theta)}{4}. 
\end{equation} 
Employing Eq.~(\ref{GreenFunction}), the second order expansion of the fermion determinant in  $\partial_\tau$ yields
\begin{eqnarray}
\label{eq:S_f_sigma_gauged}
S_2[\phi,h] & = & \frac{N}{2} {\rm Tr} \Bigl( \partial_\tau  G[h] \partial_\tau  G[h]   
-  {\dot\phi} \,G[h] \, \dot{\phi} \,G[h]     \nonumber \\
 & - & 2 i  \dot{\phi} \, G[h] \, \partial_\tau  G[h] \Bigr)
 = S_{hh} + S_{\phi\phi} + S_{\phi h} .
 \end{eqnarray}
The phase-free contribution $S_{hh}$ generates the Schwarzian term in the action~(\ref{eq:S_loc_SY}). Evaluating it along
the lines of Ref.~\cite{Bagrets-Altland-Kamenev2016} we obtain the intermediate result
\begin{equation}
S_{\rm hh} =- \frac{N  }{32 J}\, \sqrt{ \frac{\cos 2\theta }{{2\pi}}} \int\!\!\!\!\int d^2\tau
\left ( \frac{h'_1 h'_2}{|h_1-h_2|^{2}} \right)^{3/2},
\end{equation}
with the shorthand notation $h_i=h(\tau_i)$. To deal with the singularity from small
time differences, we employ the short-time Eq.~(\ref{eq:Schw_expand}) below with
$\Delta = 3/2$. In this way, the integral over short time differences $\tau_{12}\equiv \tau_1-\tau_2$
\begin{equation}
\label{eq:log}
{\cal I} = \int\limits_{ 1/J < |\tau_{12}| < m} \frac{d\tau_{12}}{|\tau_{12}|} = 2 \ln N,
\end{equation}
gives a factor logarithmic in $N$, 
and we find that in the local approximation  $S_{hh}$
turns into the Schwarzian action with the mass~(\ref{eq:m}).

In a similar manner, the phase action $S_{\phi\phi}$ assumes the nonlocal form
\begin{equation}
S_{\phi\phi} = \frac{N}{8 J}\, \sqrt{ \frac{\cos 2\theta }{{2\pi}}} \int\!\!\!\!\int d^2 \tau\, \dot\phi_1 \dot\phi_2
\left ( \frac{h'_1 h'_2}{|h_1-h_2|^{2}} \right)^{1/4} .
\end{equation} 
We once more use Eq.~(\ref{eq:Schw_expand}), now with $\Delta = 1/4$. Integration
over the time difference and focusing on the leading term (independent of $h$)  then
again produces the logarithmic factor  Eq.~(\ref{eq:log}). In this way, we  arrive at
the Coulomb phase action $S_{\phi\phi} = \tfrac K2 \int d\tau \dot\phi^2$ with $K =
32 m$. This action  describes charging effects in the unperturbed complex SYK model.
Noting that $\phi$ is canonically conjugate to the integer charge, $n$, the phase
action proportional to $K$ generates an intrinsic charging energy $E_\mathrm{GS}(n+1)
-2E_\mathrm{GS}(n)+E_\mathrm{GS}(n-1)=  K^{-1} \sim J/N$ inverse in $K$, where
$E_\mathrm{GS}(n)$ is ground state energy in $n$-particle space. This  `intrinsic'
charging energy adds to the extrinsic one, coming from the universal Hamiltonian
contribution~ $E_c \to E_c + K^{-1}$, cf. Eq.~(\ref{eq:SYKBasic}). We note that this
implies a lower limit for the actual charging energy $E_C> J/N$.

Finally, the action  $S_{\phi h}$ is parametrically small in $1/\ln N \ll 1$ and we
will not dwell on its quite technical  derivation. However, we note that it, too, is reducible to a time local form, $S_{\phi h} = - i {\cal L} \int d\tau \phi' h''/h'$,
where the logarithmic integral~(\ref{eq:log}) is not involved. This term may be
absorbed into $S_{\phi \phi}$ and $S_{h h}$ by an inessential redefinition of
parameters $K$ and $m$, justifying its omission in the action $S_0[\phi,h]$.

\emph{Winding numbers:} The quantization of charge in the isolated system is obtained
by summation over winding numbers of the $U(1)$ phase $\phi$ in imaginary time. To
explore this point within the finite temperature theory, we follow
Sachdev~\cite{Sachdev} and start from the generalization of the $T=0$ mean field
Green function to finite temperatures by choosing $\phi_T(\tau)
= - 2\pi i {\cal E} T \tau$ and $h_T(\tau) = \tan(\pi T \tau)/\pi T$ in Eq.~(\ref{GreenFunction}), 
where the parameter ${\cal E}$ is implicitly defined by $e^{2\pi{\cal E}} = \tan(\theta + \pi/4)$. This
defines the finite temperature mean field Green function
\begin{equation}
\label{eq:GT_SY}
G_T(\tau) = \mp C e^{- 2\pi \cal E T \tau} \sin(\pi/4 \mp \theta) \left(\frac{TJ}{ \sin{(\pi T |\tau|)}}\right)^{1/2},
\end{equation}
where $C=[{ (8/\pi) \cos(2\theta)}]^{-1/4}$ and the sign $\pm$ refers to ${\rm sgn}(\tau)$. The full fermion Green function, obtained from the above mean field by dressing with the phase field as it was defined in Eq.~(\ref{GreenFunction}), 
must obey the anti-periodicity condition 
$G_T(\beta/2)  =  - G_T(-\beta/2)$. Comparison with Eq.~\eqref{eq:GT_SY} shows that the phase must then satisfy 
$\phi(\beta/2) =  \phi(-\beta/2) - 2 \pi i{\cal E} + 2\pi W$, 
where $W \in \mathds{Z}$ is a winding number~\cite{Wang1996}. 

Building on this observation one can now evaluate the partition sum $Z(\beta)$ of the
complex SYK model and the correlator~$D(\tau)$. To this end, we decompose
$\phi(\tau) = \eta(\tau) + 2\pi (W - i {\cal E}) T \tau$, where $\eta(\tau)$ is
periodic, $\eta(\beta/2) = \eta(-\beta/2)$. Evaluating the Gausian integral over
$\eta(\tau)$ and using the Poisson resummation formula to rewrite the result in the
basis of quantized charge states $|n\rangle$,  the partition sum factorizes as 
${\cal Z}(\beta) ={\cal Z}_{\rm SYK}( \beta) \times {\cal Z}_C(\beta)$, 
where
\begin{equation}
\label{eq:Z_C}
{\cal Z}_C(\beta) = \sum_{n \in \mathds{Z}}e^{-\frac 12 E_c n^2/T + 2 \pi n {\cal E}},
\end{equation}
and ${\cal Z}_{\rm SYK}(\beta) \propto (mT)^{3/2} e^{2 m \pi^2 T}$ is identical to the charge neutral (Majorana) SYK model. 
The factor $e^{2 \pi n {\cal E}}$ in Eq.~(\ref{eq:Z_C}) is due to the change in the ground states
entropies for different $n$, see Ref.~\cite{Sachdev}.
In a similar manner, the Coulomb correlator $D(\tau)$  is obtained as
\begin{equation}
\label{eq:D_tau}
D(\tau) = e^{ - \frac 12 {E_C|\tau| }\left (1 - T |\tau|\right)} \times
 \frac{1}{{\cal Z}_C}\sum_{n \in \mathds{Z}} e^{- \frac 1 2 E_c (n + T \tau )^2/T + 2 \pi n {\cal E}}.
 \end{equation}
At $T \ll E_C$ the sum is dominated by $n=0$ and one recovers Eq.~(\ref{CB}). 

\emph{Cotunneling:} In the following, we derive the linear response representation of the cotunneling conductance and outline how the result motivated by scaling arguments, see Eq.~(\ref{second-order}), is obtained by a more rigorous calculation. To start with, we consider the tunneling action 
\begin{align}
\label{eq:S_T}
& S_T[{\cal A};\phi, h] = \\
& \sum_{l=R,L} \!\! g_0^{(l)} \!\! \int\!\!\!\!\int \!\! d^2 \tau e^{ - i ( {\cal  A}^{(l)}_1 + \phi_1)} g(\tau_{12})
e^{ i ( {\cal  A}^{(l)}_2 + \phi_2)} 
G_{\tau_2, \tau_1}[h], \nonumber
\end{align}
which describes the coupling of the SYK quantum dot to left (L) and right (R) leads.
Here the phase-exponentials describe the charging and un-charing of the system upon
tunneling, ${\cal  A}^{R/L}$ are source vector potentials in the right/left lead
minimally coupled to the phase, and $g(\tau) = - \pi T/\sin(\pi T \tau)$ are the
Fermi liquid lead Greens functions. To 1st order in $g_0$, this action  describes the
sequential tunneling of electrons from the leads to the dot and back. As outlined in
the main text, the corresponding contribution to the conductance is suppressed as
$e^{-E_C/T}$ at low temperatures $T \ll E_C$.  The cotunneling contribution to $g(T)$
is obtained from the 2nd order cumulant expansion in $g_0$, which after
re-exponentiation is described by the action $S_2[{\cal A}] = - \tfrac 12 \langle
\left( S_T[{\cal A}; \phi,h]\right)^2 \rangle_{\phi,h}$, where the average is
performed over the action~$S_0[\phi,h]$. In more explicit terms $S_2[{\cal A}]$
reads
\begin{align}
& S_2[{\cal A}] = - g_0^{(L)}g_0^{(R)}\! \int \! d^4\tau
e^{i ( {\cal  A}^{(L)}_2 - {\cal  A}^{(L)}_1 + {\cal  A}^{(R)}_4 - {\cal  A}^{(R)}_3 ) } \\ \nonumber 
&g(\tau_{12}) g(\tau_{34}) \langle e^{ i (  \phi_2 - \phi_1  + \phi_4 - \phi_3)}\rangle_\phi 
  \left\langle G_{\tau_1,\tau_2}[h] G_{\tau_3,\tau_4}[h] \right\rangle_h\! . 
\end{align}
Various contributions to the cotunneling channel are now distinguished via the
ordering  of the time arguments $\tau_i$. For instance, the sequence $\tau_1 > \tau_4
> \tau_2 > \tau_3$ describes the exiting of a dot electron  to the right lead
at time $\tau_3$ followed by the entering of a left lead electron onto the dot at
time $\tau_2$. This process is accompanied by the creation of a particle/hole excitation    via tunneling from/to
the right/left lead at times $\tau_4$/$\tau_1$, respectively. 
The Feynman diagram representing these processes is depicted in the inset of
Fig.~\ref{fig:InelasticCotunneling}, where the internal bubbles symbolize the internal interaction processes due to the presence of $\hat H_\mathrm{SYK}$ (represented by $h$-fluctuations in the low energy limit) and the dashed lines the correlation via $\phi$.
The latter enter the 
time-resolved probability of the  virtual cotunneling  excitation via the factor
\begin{equation} 
\langle n | e^{  - i\hat\phi(\tau_1) + i\hat\phi(\tau_2)  - i\hat\phi(\tau_3) + i\hat\phi(\tau_4) } | n \rangle \!=\! 
e^{- \tfrac 12 E_c ( \tau_{14} + \tau_{23})},
\end{equation}
where $\tau_{kl} = \tau_k - \tau_l$. We are here working in a representation of
 integer charge states with $e^{\pm i \hat \phi} | n \rangle = | n \pm 1\rangle$
 where $|n\rangle$ is the ground state of the dot and the $U(1)$ act as Heisenberg
 operators, $\phi(\tau_k) \to \hat \phi(\tau_k) - 2 i \pi {\cal E} T \tau_k$. 
 Integrating over time differences $\tau_{14}, \tau_{23}$  and summing over all
 different time orderings, one arrives at the intermediate result
\begin{equation}
\label{eq:Ic}
S_2[{\cal A}] =  \frac{g_0^{(L)}g_0^{(R)}}{E_C^2}  \sum_{\sigma = \pm } \int  d^2\tau e^{i \sigma ({\cal A}_1 - {\cal A}_2 )} 
{\cal K}(\tau_{12}),
\end{equation}
where ${\cal A} = {\cal A}^{(R)} - {\cal A}^{(L)}$ and we have introduced the response kernel
\begin{equation}
{\cal K}_{\rm it}(\tau) = - g (\tau_{12}) g(\tau_{21}) \left\langle G_{\tau_1,\tau_2}[h] G_{\tau_2,\tau_1}[h] \right\rangle_h.
\end{equation}
After the average over $h$, the latter becomes a function of $\tau= \tau_1 - \tau_2$ and can be continued to real times as
\begin{equation}
\label{eq:K_gl}
{\cal K}_{\rm it}^{\gtrless}(t) = i {\cal K}_{\rm it}(\tau \gtrless 0)\Bigl|_{\tau \to it \pm 0}.
\end{equation}
 The linear response DC cotunneling
conductance is then obtained from  the action~(\ref{eq:Ic}) in the usual manner by two-fold differentiation in the source vector potential, division by a real frequency parameter, $\omega$, and the taking of a zero frequency limit. Substituting factors, this gives 
\begin{equation}
\label{eq:git_K}
g_{\rm it}(T) =   \frac{g_0^{(L)}g_0^{(R)}}{E_C^2}\,  \partial_\omega {\rm Im}\, {\cal K}_{\rm it}(\omega)\Bigl|_{\omega=0},
\end{equation} 
where ${\rm Im} \,{\cal K}_{\rm it}(\omega) =  \tfrac i2 \left( {\cal K}_{\rm it}^>(\omega) - {\cal K}_{\rm it}^<(\omega) \right ) $.

From the general relation~(\ref{eq:git_K}) different cases of physical interest can
be analyzed: For large temperatures,  $m^{-1} \ll T\ll E_C$, the reparameterization
degree of freedom,  $h$, is locked to $h_T(\tau)$. In this case, $\left\langle
G_{12}[h] G_{21} [h] \right\rangle_h \simeq G_T(\tau) G_T(-\tau)$, and the kernel
${\cal K}$ reads
\begin{equation}
{\cal K}_{\rm it}(\tau) \propto - \frac{T^3J}{\sin^3(\pi T |\tau|)}, \quad m^{-1}< T< J,E_C.
\end{equation}
With the help of Eqs.~(\ref{eq:git_K}) and (\ref{eq:K_gl}) the cotunneling
conductance then becomes
\begin{equation}
\label{eq:git1}
g_{\rm it}(T) \propto   - \frac{g_0^2 T^3J}{E_C^2} {\rm Re } \int\limits_{-\infty}^{+\infty} 
\frac{t dt}{\sinh^3(\pi T (t-i 0))} = \frac{g_0^2 \,TJ}{4E_C^2}.
\end{equation} 
The singularity at $t=0$ in this integral can be removed by 
shifting the integration contour to the lower complex half-plane as $t= u - i \beta/2$ with $u \in \mathds{R}$.

In the opposite regime of strong reparametrization fluctuations, $T \ll m^{-1}$, one finds~\cite{Bagrets-Altland-Kamenev2016} $\left\langle G_{12}[h] G_{21} [h] \right\rangle_h \stackrel{T\to 0}\sim  {J m^{1/2}}/{ \tau^{3/2}}$. Generalizing this result to finite $T$ along the lines of Refs.~\cite{bagrets2017power, Mertens2017}
gives the response kernel
\begin{equation}
{\cal K}_{\rm it}(\tau) \propto - \frac{J(mT)^{1/2}}{ \sin^2(\pi T \tau)} \times
\frac{1}{|\tau|^{3/2}(\beta-|\tau|)^{3/2}}, \quad T< m^{-1}.
\end{equation}
Therefore the cotunneling conductance in this regime becomes
\begin{eqnarray}
\label{eq:git2}
g_{\rm it}(T) &\propto& - \frac{g_0^2 J(mT)^{1/2}}{E_C^2} {\rm Re } \int\limits_{-\infty}^{+\infty} 
\frac{t dt}{\sinh^2(\pi T (t-i 0))}  \\
&\times& \frac{1} {(it + 0)^{3/2}(\beta - it - 0)^{3/2}} \propto \frac{g_0^2 J  m^{1/2} T^{3/2}}{E_C^2}, \nonumber
\end{eqnarray}
where the integral can be evaluated using a shift of the integration contour similar to Eq.~(\ref{eq:git1}) above. 
The results~(\ref{eq:git1}) and (\ref{eq:git2}), summarized by Eq.~(8) of the main text
and valid in the SYK phase, should be contrasted to the Fermi liquid case
where ${\cal K}_{\rm FL}(\tau) \propto - {T^4}/{\sin^4(\pi T \tau)}$ and $g_{\rm it}(T) \propto g_0^2 T^2/E_C^2$
in agreement with Ref.~\cite{AverinNazarov}.

\emph{Tunneling conductance:} The direct tunneling contribution to conductance (see Fig.~1) can be evaluated along 
the same lines as above. For that, we introduce the response kernel 
\begin{equation}
{\cal K}_{\rm dt}(\tau) = g(\tau_{21}) \langle G_{\tau_1,\tau_2}[h]\rangle_h D(\tau_{12}),
\end{equation} 
of the lowest order tunneling action~(\ref{eq:S_T}). The direct tunneling conductance satisfies
$1/g_{\rm dt} = 1/g^{(L)} + 1/g^{(R)}$, with $g^{(l)}$ being direct tunneling conductances from the $l$-th lead to the dot. 
Hence the analog of Eq.~(\ref{eq:git_K}) becomes
\begin{equation}
\label{eq:gdt_K}
g_{\rm dt}(T) =   \frac{g_0^{(L)}g_0^{(R)}}{g_0^{(L)} + g_0^{(R)}}\, 
 \partial_\omega {\rm Im}\, {\cal K}_{\rm dt}(\omega)\Bigl|_{\omega=0}.
\end{equation} 
To simplify the further analysis we consider a particle-hole symmetric point ${\cal E} = 0$ (the results for general ${\cal E}$ are qualitatively the same). In this case the Coulomb correlator~(\ref{eq:D_tau}) can be written in  closed form as
\begin{equation}
D(\tau) = \frac{\vartheta_3( \tfrac i2 E_c \tau, e^{-E_c/2T})e^{- \tfrac 12 E_c|\tau|}}{\vartheta_3( 0, e^{-E_c/2T})},
\end{equation}
where $\vartheta_3(u,q) $ is the Jacobi theta function. Also, since $E_c \gg m^{-1}$, it is sufficient to analyze the limit of high temperatures, $T \gg m^{-1}$ (at smaller $T$ the conductance is dominated by cotunneling anyway).
Hence we take the response kernel to be
\begin{equation}
{\cal K}_{\rm dt}(\tau) \propto - \frac{T^{3/2} J^{1/2}  D(\tau)}{\sin^{3/2} (\pi T |\tau|)}, \quad T > m^{-1}.
\end{equation}
Analytically continuing ${\cal K}_{\rm dt}(\tau)$ to real times, cf.~Eq.~(\ref{eq:K_gl}), and using the Kubo formula~(\ref{eq:gdt_K}),
the direct tunneling conductance reads
\begin{eqnarray}
g_{\rm dt}(T) &\propto& \frac{g_0 T^{3/2} J^{1/2} }{\vartheta_3( 0, e^{-E_c/2T})} \\
 &\times& {\rm Im} \int\limits_{-\infty}^{+\infty}
\frac{ \vartheta_3( \tfrac 12 E_c t, e^{-E_c/2T}) e^{-\tfrac i2 E_c t}\,t dt}{\sin^{3/2} (\pi T ( i t + 0))}. \nonumber
\end{eqnarray}
For the practical evaluation of this integral one
shifts the integration contour to the lower complex half-plane as $t= u - i \beta/2$ with $u \in \mathds{R}$.
On evaluating it numerically we get the result in the form $g(T) \propto g_0 (J/T)^{1/2} f(T/E_c)$, where
$f(x)$ is the scaling function with the properties $f(+\infty) = 1$ and $f(x) \sim e^{-1/x}$ at $x \to 0$ (see Fig.~1).

\emph{RG analysis:} We here discuss how the stability of the SYK phase to single particle perturbations of moderate strengths is established via an RG procedure. Following the standard RG recipe,  we decompose  reparameterization fluctuations onto 'fast' and 'slow' as  $h(\tau) =
f(s(\tau)) \equiv (f \circ s)(\tau)$, where
 $f$ and $s$ are fluctuations in the frequency  range
$[\Lambda, J]$ and $[0,\Lambda]$, and $\Lambda$ is a running cutoff energy. We then integrate out the fast modes $f(s)$,
and rescale time $\tau\to \tau J/\Lambda$ to restore the UV cutoff $\Lambda\to J$.  
Consider first the case $m^{-1}< \Lambda<J$, where the reparameterization fluctuations are suppressed. 
The RG flow is then governed by the `engineering' dimensions of the running constants, resulting in:
\begin{equation}
						\label{eq:RG-bare}
\frac{d\ln m}{dl}=-1;\quad\quad \frac{d\ln w}{dl} = +1,
\end{equation}
where $l=\ln(J/\Lambda)$. For $T>J/N$ this flow should be terminated when either
$\Lambda$ reaches $T$, or $W(l)\sim \sqrt{w(l)}$ reaches the UV cutoff $J$. This
defines the temperature scale $T_\mathrm{FL}\propto W^2/J$,  separating the high temperature
$g(T)\propto 1/\sqrt{T}$ regime, see Eq.~(\ref{first-order}),  
and low temperature FL regime, where $g(T)\propto T^2$, \cite{AverinNazarov}.  

We now turn to the regime of strong reparameterization fluctuations. When 
$\Lambda$ reaches $J/N$, $m(l)=m(0)e^{-l}$ reaches the inverse UV cutoff $m(l)\approx
1/J$. To proceed\cite{Altland2019}, we employ the Schwarzian chain
rule
\begin{equation}
					\label{eq:chain}
\{f\circ s,\tau\}=(s')^2\{f,s\}+\{s,\tau\}, 
\end{equation}
to obtain the action: $S_0[f \circ s] = S_0^{\rm fast}[f,s] + S_0[s]$,
where 
the 'fast' Schwarzian action has a time-dependent mass
$m(s)\equiv  m s^{a\prime }$.
At  lowest order in $w$ one needs to average the  action
$S_{W}[f\circ s]$ over the fast fluctuations.
A straightforward application of the chain rule to  the Green functions (\ref{GreenFunction}) shows that  
\begin{align}
\label{eq:GF-chain} 
G_{\tau_1,\tau_2}[f\circ s]=G_{s_1,s_2}[f](s'_1s'_2)^{1/4}, 
\end{align}
so that $\langle S_{\mathrm{W}}[f\circ s]\rangle_f
\propto  \langle \big( G_{s_1,s_2}[f]\big)^2 \rangle_{f} $. This can be evaluated with the help of exact results \cite{Bagrets-Altland-Kamenev2016,Mertens2017} for the 4-point propagator of the Schwarzian theory.
Following Ref.~\cite{Altland2019}, one finds for the asymptotic expressions ($s_{12}\equiv s_1-s_2$): 
\begin{equation}
						\label{eq:GF-asymptotic}
\langle \big(G_{s_1,s_2}[f]\big)^2\rangle_f  \simeq \begin{cases}
|s_{12}|^{-1}, &  \!\!\!\!\!\!\! s_{12}< m;\\   
\!\bigl(m(s_1)m(s_2)\bigr)^{1/4}  |s_{12}|^{-3/2}, & \\
\sqrt{m\Lambda} |s_{12}|^{-1}, &\!\!\!\!\!\!\! \Lambda^{-1} < s_{12},   
\end{cases} 
\end{equation}
where the middle line is for $ m < s_{12}<\Lambda^{-1} $.
This equation implies that the double time integral  in the averaged
action $\langle S_{\mathrm{W}}[f\circ s]\rangle_f\equiv S_{\mathrm{int}}+S_{\mathrm{long}}$ gets different contributions from intermediate ($m <\tau_{12}< \Lambda^{-1}$) and long time differences ($\tau_{12}>\Lambda^{-1}$). The former is evaluated with the help of the Taylor expansion ($\tau = (\tau_1 + \tau_2)/2$)
\begin{align}
                            \label{eq:Schw_expand}
\left( \frac{s'_1 s'_2}{[s_1-s_2]^2}  \right)^\Delta \approx
\frac{1}{[\tau_1 - \tau_2]^{2\Delta}} + \frac{\Delta}{6} 
\frac{\{s(\tau),\tau\} }{[\tau_1 - \tau_2]^{2\Delta-2}}  + \dots,  
\end{align}
with $\Delta=3/4$. Integrating the last term in this expression over $\tau_1-\tau_2$ over the intermediate window in Eq.~(\ref{eq:GF-asymptotic}), one finds correction to the $m$-coupling of the form $\delta m = \frac{\Delta}{6} \frac{2}{3}w m^2 (m\Lambda)^{-3/2}$. Introducing now the logarithmic scale for the running cutoff $l=-\ln(m\Lambda)$ and   
 rescaling time to retain the value of the cutoff, $\Lambda$, we find $m$-renormalization 
$m\to m(l)\equiv e^{-l}(m+
\frac{1}{12} w m^2 e^{3l/2})$. The  integration over large time differences conserves the form of the  $W$-action, but changes the coupling constant as $w\to w(l) =e^l w(m\Lambda)^{1/2} = e^lwe^{-l/2}$. Differentiating these expressions over $l$ at $l=0$, one finds  
\begin{equation}
 						\label{eq:RG-1order1}
\frac{d\ln m}{dl} =-1 + \frac{1}{24} w m;\quad\quad \frac{d\ln w}{dl} = {1\over 2},
\end{equation} 
which is Eq.~(\ref{eq:RG-1order}) of the main text. Notice  the change of the scaling dimension of the coupling $w$ between weak, Eq.~(\ref{eq:RG-bare}), and strong, Eq.~(\ref{eq:RG-1order1}), fluctuation regimes. It remains a relevant perturbation in
both of them. However the dimensionless coupling constant $\lambda\equiv  wm$ obeys $d\lambda/dl =0$ for $m\gg 1/J$ and  $d\lambda/dl =\lambda/24-1/2$ for $m\lesssim 1/J$. Thus $\lambda_c=12$ is the critical value between the two regimes.  

\bibliography{SYK}

\end{document}